\documentclass[%
 reprint,
superscriptaddress,
 amsmath,amssymb,
 aps,
]{revtex4-2}

\usepackage{graphicx}
\usepackage{dcolumn}
\usepackage{bm}

\newcommand{\ket}[1]{\vert#1\rangle}
\newcommand{\bra}[1]{\langle#1\vert}
\newcommand{\kb}[2]{\vert#1\rangle\!\langle#2\vert}
\newcommand{\bk}[2]{\langle#1\vert#2\rangle}
\newcommand{\diff}{\mathrm{d}}

\newcommand{\vk}{\mathbf{k}}
\newcommand{\vu}{\mathbf{u}}
\newcommand{\vz}{\mathbf{z}}

\newcommand{\mean}[1]{\mathcal{M}\left[#1\right]}
\newcommand{\vmu}{\boldsymbol{\mu}}
\newcommand{\vEps}{\boldsymbol{\mathcal{E}}}
\newcommand{\Eps}{\mathcal{E}}
\newcommand{\integer}{\mathbb{N}}

\newcommand{\abs}[1]{\left|#1\right|}

\newcommand{\ptr}[2]{\mathrm{tr}_{#2}\left\{#1\right\}}

\begin{document}


\title{Linear optical properties of organic microcavity polaritons with non-Markovian Quantum State Diffusion}
\author{Timo Leppälä}%
\affiliation{Department of Physics and Astronomy, University of Turku, Turku, Finland}
\author{Ahmed Gaber Abdelmagid}
\affiliation{Department of Mechanical and Materials Engineering, University of Turku, Turku, Finland}
\author{Hassan A. Qureshi}
\affiliation{Department of Mechanical and Materials Engineering, University of Turku, Turku, Finland}
\author{Konstantinos S. Daskalakis}
\affiliation{Department of Mechanical and Materials Engineering, University of Turku, Turku, Finland}
\author{Kimmo Luoma}%
\affiliation{Department of Physics and Astronomy, University of Turku, Turku, Finland}
\email{ktluom@utu.fi}

\date{\today}

\begin{abstract}
    Hybridisation of the cavity modes and the excitons to polariton states together with the coupling to the vibrational modes determine the linear optical properties of organic semiconductors in microcavities. In this article we compute the refractive index for such system using the Holstein-Tavis-Cummings model and determine then the linear optical properties using the transfer matrix method.  We first extract the parameters for the exciton in our model from fitting to experimentally measured absorption of a 2,7-bis [9,9-di(4-methylphenyl)-fluoren-2-yl]- 9,9-di(4-methylphenyl) fluorene (TDAF) molecular thin film. Then we compute the reflectivity of such a thin film in a metal clad microcavity system by including the dispersive microcavity mode to the model. We compute susceptibility of the model systems evolving just a single state vector by using the non-Markovian Quantum State Diffusion.
    The computed location and height of the lower and upper polaritons agree with the experiment within the estimated errorbars for small angles ($\leq 30^\circ$). For larger angles the location of the polariton resonances are within the estimated error.
\end{abstract}

\maketitle
\section{Introduction}
\begin{figure*}[t]
  \includegraphics[width=0.99\textwidth]{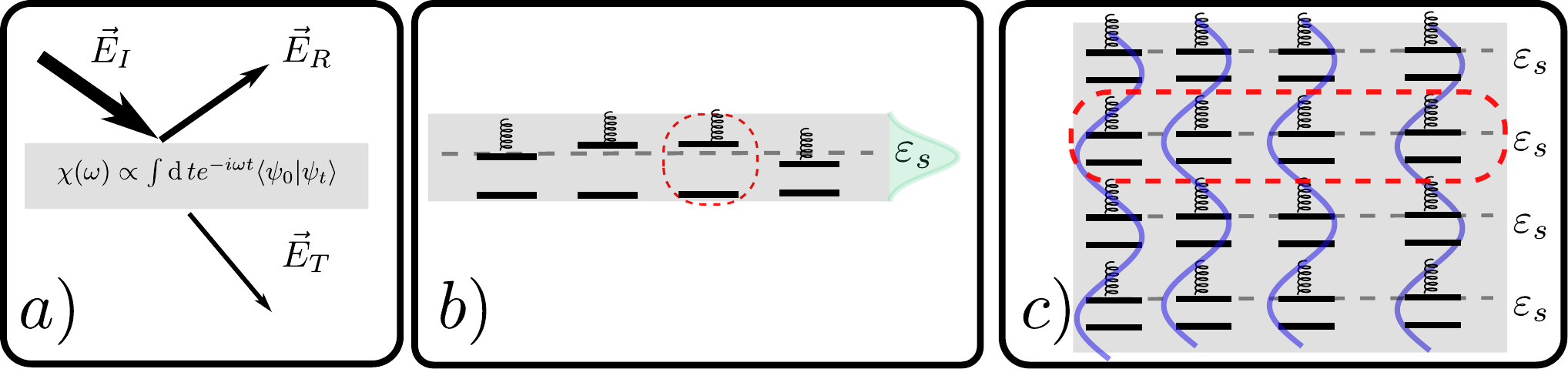}
  \caption{\label{fig:schematic} \textbf{a)} We model the linear optical properties of a slab of material
  in terms of the dielectric function $\varepsilon$, which can be computed from the susceptibility $\chi$. The susceptibility is computed from a quantum mechanical model evolving just a single state vector. This leads to a computationally efficient scheme in contrast to approaches where density matrix evolution is needed.\textbf{b)} Susceptibility of a thin film can be modeled as the susceptibility of a single quantum absorber (red box) multiplied with the the number of such absorbers, when spatial disorder can be neglected. We include energetic disorder as indicated by the distribution of exciton energies $\varepsilon_s$, and a coupling to vibration models when the system is excited indicated by the spring.\textbf{c)} In the case of microcavity polaritons we consider a slice of emitters in the $z$-direction (red box), whereas the cavity mirrors are located at $x=\pm L/2$ (top and bottom). }
\end{figure*}
{The localized nature of molecular excitons imparts two crucial attributes: a large binding energy and a dense population of excitons compared to the photonic density of states.
The former feature promotes strong light-matter coupling at room temperature, making possible the macroscopic study of Bose-Einstein condensation and superfluidity at elevated temperatures~\cite{Keeling2020,Tang2021, Moilanen2021, Castellanos2023}. The latter feature enabled the observation of ultrastrong coupling~\cite{FriskKockum2019} and single photon nonlinearity~\cite{Zasedatelev2021}. Recently, the concept of polariton chemistry was introduced, which promises to reshape the energy landscape of molecular systems, exerting control over their photochemical and photophysical processes~\cite{Sanvitto2016,Hertzog2019,Yuen-Zhou2020,Garcia-Vidal2021,Bhuyan2023}. In optical microcavities filled with molecular absorbers, polariton modes are the eigenstates resulting from the strong coupling between the optical modes and the exciton resonances. They can be observed experimentally as the avoided crossing of the bare exciton and microcavity photon dispersion through reflectivity measurements~\cite{Torma2015}. 

An interesting implication of strong coupling at non-zero temperature is that the exciton might also strongly couple to the surrounding molecular vibrations which can be the catalyst for complex relaxation dynamics of polaritons~\cite{Hulkko2021}. Currently, there is significant research being conducted to explore the impact of strong collective light-matter interactions on the intersystem crossing (ISC) and reverse intersystem crossing (RISC) rates~\cite{Stranius2018, Martinez-Martinez2018, Berghuis2019, Eizner2019, Polak2020, Yu2021,Mukherjee2023, abdelmagid2023}. Rabi splitting is proportional to the square root of the molecular density that share the microcavity photon ($\sqrt{N/V}$). To achieve strong coupling in planar microcavities with highly delocalized photonic mode, a large number of molecules $N$, in the order of 10$^6$ is needed. However, perturbative quantum mechanical calculations predict that possible polaritonic contributions to the  RISC or ISC processes, for example, scale with inverse of this number~\cite{Tichauer2021,Martinez-Martinez2019,Sanchez-Barquilla2022,Miwa2023}. 
}

The purpose of this article is to introduce a new method for investigating the polariton dynamics and pave the way for gaining new insights into the so called "large number of molecules $N$" 
problem. In this work we take the first steps of applying non-Markovian Quantum State Diffusion (NMQSD) for computing the linear optical properties of organic microavity polaritons. While a possible drawback of this approach is that it is stochastic and typically requires a large number of trajectories for computing expectation values of observables, 
for computing linear optical properties using just one trajectory is enough~\cite{Roden_2009,roden_2011,Ritschel_2015}.

In Fig.~\ref{fig:schematic} we present a graphical summary of this article. \textbf{a)} We use NMQSD to compute the susceptiblity of the model systems evolving just a single state vector.  From the susceptibility we obtain the refractive index which we use to model the experimental situations studied in this article. \textbf{b)} We compute the absorption of a thin film of TDAF molecules. By fitting the model to the experiment we obtain the parameters of the exciton. \textbf{c)} Lastly, we investigate a situation where the TDAF molecules are placed within the microcavity and we compute the reflectivity of such a system.

The structure of this article is the following. In Sec.~\ref{sec:linear_response} we introduce the linear response theory and how the susceptibility and refractive index can be computed. In Sec.~\ref{sec:non-mark-quant} we introduce the NMQSD method. Furthermore, we show how the susceptibility can be computed using the NMQSD approach. Then in Sec.~\ref{sec:thin_film} we fit the model parameters to the single molecule data obtained from experimental measurements and density functional calculations. We will focus on 2,7-bis [9,9-di(4-methylphenyl)-fluoren-2-yl]- 9,9-di(4-methylphenyl) fluorene (TDAF) as it is a model system for strong light-matter studies~\cite{Daskalakis2018, palo2023, abdelmagid2023}. In Sec.~\ref{sec:polariton_reflectivity} we construct the model for microcavity polaritons using the same TDAF molecule. We discuss in detail how the susceptibility can be computed in this case when the dispersive microcavity mode is also included. We also present the results of the theoretical calculations. In Sec.~\ref{sec:conclusions_and_outlook} we present our conclusions and outlook. Experimental details are presented in Sec.~\ref{sec:methods}
\section{Linear response}\label{sec:linear_response}
\paragraph{Linear optics}In linear materials the polarization field is proportional to the applied electric field
\begin{align}
 \mathbf{P}(t,z)=\int\limits_{0}^t\diff s\int\diff z'\,\varepsilon_0\chi(t-s,z-z')\vEps(s,z'),
\end{align}
where $\varepsilon_0$ is the vacuum permittivity and $\chi$ is the susceptibility. We set $\hbar=1$ in all subsequent equations. Due to causality polarization can depend only on the fields applied 
on earlier times and may have non-local spatial dependency~\cite{kavokin2007microcavities}. In general, $\chi$ is a second rank tensor. In this work we focus only on situations where the polarization is aligned with the applied electric
field making $\chi$ a scalar. By using the convolution theorem and after dropping the vector notation this relation is
\begin{align}\label{eq:linear_polarization}
    P(\omega,k_z) = \varepsilon_0\chi(\omega,k_z)\Eps(\omega,k_z),
\end{align}
where $\omega$ is the angular frequency and $k_z$ is the $z$-component of the wave vector. The dielectric function (or relative permittivity) is  
\begin{align}
    \varepsilon(\omega,k_z)=1+\chi(\omega,k_z).
\end{align}
The dielectric function determines the refractive index by $n=\sqrt{\varepsilon}$.  Once the refractive index is known, the linear optical properties of planar systems can be computed using the transfer matrix method (TMM)\cite{Panzarini1999}.

\paragraph{Dipole density}Polarization corresponds to the dipole density of the medium, which can be computed from a microscopical model.  We consider a situation where weak classical field is used to probe a quantum system.  The interaction term between the field and the matter is taken to be 
\begin{align}
H_F(t) = -\sum_m\Eps(z_m,t)\cdot\mu_m,
\end{align}
where $z_m$ is the location of the dipole and 
$mu_m$ is the dipole operator. The other degrees of freedom are described by a time independent Hamilton operator $H$. The dynamics generated by $H$ is given by the unitary operator
\begin{align}
    U(t) = e^{-i H t}.
\end{align}
The linear response can be computed from the dipole correlation function $M(t)$~\cite{may2011charge,Roden_2009,roden_2011}. 
The susceptibility of the system can be computed as the Fourier transform of the dipole correlation function $M(t)$
\begin{align}\label{eq:linear_susceptibility}
    \chi(\omega,k) = \int_{0}^\infty\diff t\, e^{i\omega t}M(t,k_z),
\end{align}
where 
\begin{align}\label{eq:dipole_correlation}
    M(t,k_z) = \sum_{m,n}e^{ik_z(z_m-z_n)}\bra{\Psi_G}\mu_n U(t)\mu_m\ket{\Psi_G},
\end{align}
and $\ket{\Psi_G}$ is the ground state of the Hamiltonian $H$. We assume that the systems we investigate do not have permanent dipole moment. Under this assumption the dipole correlation function is obtained from perturbation theory and keeping only the positive energy terms and the terms linearly proportional to the applied field~\cite{haug2004quantum,klingshirn1997semiconductor}. Generalization to anisotropic cases is straightforward. Using Eqs.~\eqref{eq:linear_polarization},\eqref{eq:linear_susceptibility} gives the polarization density of the system. The macroscopic polarization is obtained by multiplying the polarization density with the sample volume. In this work we neglect any spatial disorder.
Finite temperature results can be computed using the ground state initial condition in the NMQSD approach as we show later.
\section{Non-Markovian Quantum State Diffusion}\label{sec:non-mark-quant}
\paragraph{General theory}
The aim of the NMQSD approach is to solve the time evolution of the
full Schrödinger equation for the open system and the environment~\cite{Diosi1998,Hartmann_2017,Megier_2018}. A typical
model consists of an open system with Hamiltonian
$H_S$ and a coupling operator $L$. These operators are arbitrary at this point.
The environment is assumed to consist of quantum harmonic oscillators with a
Hamiltonian
\begin{align}\label{eq:H_E_gen}
  H_E = \sum_\lambda \omega_\lambda b_\lambda^\dagger b_\lambda,
\end{align}
where $[b_\lambda,b_{\lambda'}^\dagger] = \delta_{\lambda,\lambda'}$. 
Generalization to spin baths is possible~\cite{Link_2023}.
The interaction is taken to be linear in the coupling operator $L$ of the open system
and the creation and annihilation operators of the environment
\begin{align}
  H_I =  \sum_\lambda g_\lambda L a_\lambda^\dagger + g_\lambda^* L^\dagger a_\lambda.
\end{align}
The initial state of the bath is the thermal state $\rho_\beta$ and the system
and the bath are initially uncorrelated.
In the interaction picture with respect to (\ref{eq:H_E_gen}) the Schrödinger
equation is
\begin{align}\label{eq:Schroedinger_eq_gen}
  \frac{\diff}{\diff t}\ket{\Psi_t} = -i H_S\ket{\Psi_t}+\sum_\lambda \left(g_\lambda L
  a_\lambda^\dagger e^{i\omega_\lambda t} + h.c.\right)\ket{\Psi_t}
\end{align}
The finite temperature NMQSD equation corresponding to the Schrödinger equation~(\ref{eq:Schroedinger_eq_gen})
is~\cite{Diosi1998}
\begin{align}\label{eq:NMQSD_gen}
  \frac{\diff}{\diff t}\ket{\psi_t(\vz^*)}
  =&-i(H_S+V(t))\ket{\psi_t(\vz^*)}+z_t^*L\ket{\psi_t(\vz^*)}\notag\\
  &-L \int_0^t\diff s\, \alpha(t-s)\frac{\delta}{\delta z_s^*}\ket{\psi_t(\vz^*)},
\end{align}
where $z_t^*$ is a zero mean Gaussian stochastic process with correlations
\begin{align}\label{eq:noise_correlations}
  \mean{z_tz_s^*}=\alpha(t-s),\quad \mean{z_tz_s}=0.
\end{align}
The hermitian autocorrelation of the process corresponds to the 
zero temperature bath correlation function (BCF)
\begin{align}\label{eq:bcf_zero_temp}
    \alpha(t-s) = \int_0^t\diff s\, J(\omega)e^{-i\omega(t-s)},
\end{align}
We have introduced the spectral density $J(\omega)=\sum_\lambda\abs{g_\lambda}^2\delta(\omega_\lambda-\omega)$. which controls the properties of the environment. 

Finite temperature is incorporated by a "stochastic potential" $V(t)$ 
\begin{align}
    V(t) = L\eta_t+L^\dagger\eta_t^*,
\end{align}
where $\eta_t$ is a zero mean Gaussian stochastic process with correlations~\cite{Goetsch_1996,Hartmann_2017}
\begin{align}
    \mean{\eta_t\eta_s^*} = \sum_\lambda n_\lambda \abs{g_\lambda}^2 e^{-i\omega_\lambda (t-s)},
\end{align}
and $n_\lambda=(e^{\beta\omega_\lambda}-1)^{-1}$ is the thermal photon number. 
The NMQSD equation is a stochastic differential equation containing the Hamiltonian term, stochastic driving term and a memory term with a functional integral. The states $\ket{\psi_t(\vz^*)}$ are analytical functionals of the noise process $z_t^*=-i\sum_\lambda g_\lambda z_\lambda^*e^{-i\omega t}$, where $z_\lambda^*$ are the labels of the Bargmann coherent states of the environment~\cite{Megier_2018}. By construction the exact open system is recovered by averaging over the stochastic trajectories
\begin{align}\label{eq:reduced_state}
  \rho(t) = \ptr{\kb{\Psi_t}{\Psi_t}}{E} = \mean{\kb{\psi_t(\vz^*)}{\psi_t(\vz^*)}}.
\end{align}
The challenging part in using NMQSD is the occurence of the functional derivative. In recent years a powerful hierarchy of pure states (HOPS) approach has emerged as a general numerical approach to solve the NMQSD equations~\cite{Suess2014}. However, in this work we use a different approach. Namely, we  approximate the functional derivative with the following expression
\begin{align}\label{eq:Redfield}
    \frac{\delta}{\delta z_s^*}\ket{\psi_t(\vz^*)}
    =e^{-iH_S(t-s)}Le^{iH_S(t-s)}\ket{\psi_t(\vz^*)}.
\end{align}
This approximation is obtained from the general HOPS approach by truncating the hierarchy after the first level and corresponds to a weak coupling approximation (we neglect any terms higher than second order in the coupling strength)~\cite{Yu_1999,Suess2014,Hartmann_2017}.

\paragraph{Susceptibility using NMQSD.}
 Using the NMQSD we can compute the susceptibility of the system evolving pure states only. In case the coupling operator is hermitian we need to evolve only one pure state, otherwise we need to average over the thermal noise~\cite{Ritschel_2015}. In the case that we have multiple transition dipole moments we 
 need to extend the NMQSD to many systems which all couple to their individual environments. This simply means that each subsystem has their own 
 coupling operator $L_m$, noise term $z_{t,m}^*$ and bath correlation function $\alpha_m(t)$.
 We assume that the Hamiltonian is such that the global ground state is a product from the system ground state and the vacuum of the bath $\ket{\Psi_G}=\ket{g_1,0}\ket{g_2,0}\cdots\ket{g_M,0}$, where we consider $M$ systems and $\ket{0}$ is the vacuum state of the environment of the respective open system.
In the NMQSD approach the dipole correlation function $M(t)$ can be computed from the time-evolution~\cite{Roden_2009,roden_2011,Ritschel_2015}
\begin{align}\label{eq:dipole_correlation_NMQSD}
  M(t) = {\mu_{tot}^2}\bk{\psi_0}{\psi_t(\vz^*=0)},
\end{align}
where $\ket{\psi_0}$ is the initial state for the NMQSD evolution and
$\ket{\psi_t(\vz^*=0)}$ is the solution to the NMQSD equation where the driving
noise is set to zero, i.e. $z_t^*=0$. The initial condition for the evolution is chosen as
\begin{align}\label{eq:initial_condition}
  \ket{\psi_0} = \frac{1}{\mu_{tot}}\sum_m e^{ik_zz_m}\mu_m\ket{g},\quad
  \mu_{tot}= \sqrt{\sum_m\abs{\mu_m}^2}.
\end{align}
If the coupling operator is hermitian, we can replace the stochastic potential with the finite temperature bath correlation function
\begin{align}\label{eq:BCF_finite_temp}
    \alpha(t)=\int\limits_0^\infty\diff\omega\, J(\omega)\left(\coth\left(\frac{\beta\omega}{2}\right)\cos\left(\omega t\right)-i\sin\left(\omega t\right)\right).
\end{align}
Otherwise we need to average over different realizations of the thermal noise~\cite{Roden_2009}.
\section{Molecular thin film}\label{sec:thin_film}
\begin{figure}
  \includegraphics[width=0.49\textwidth]{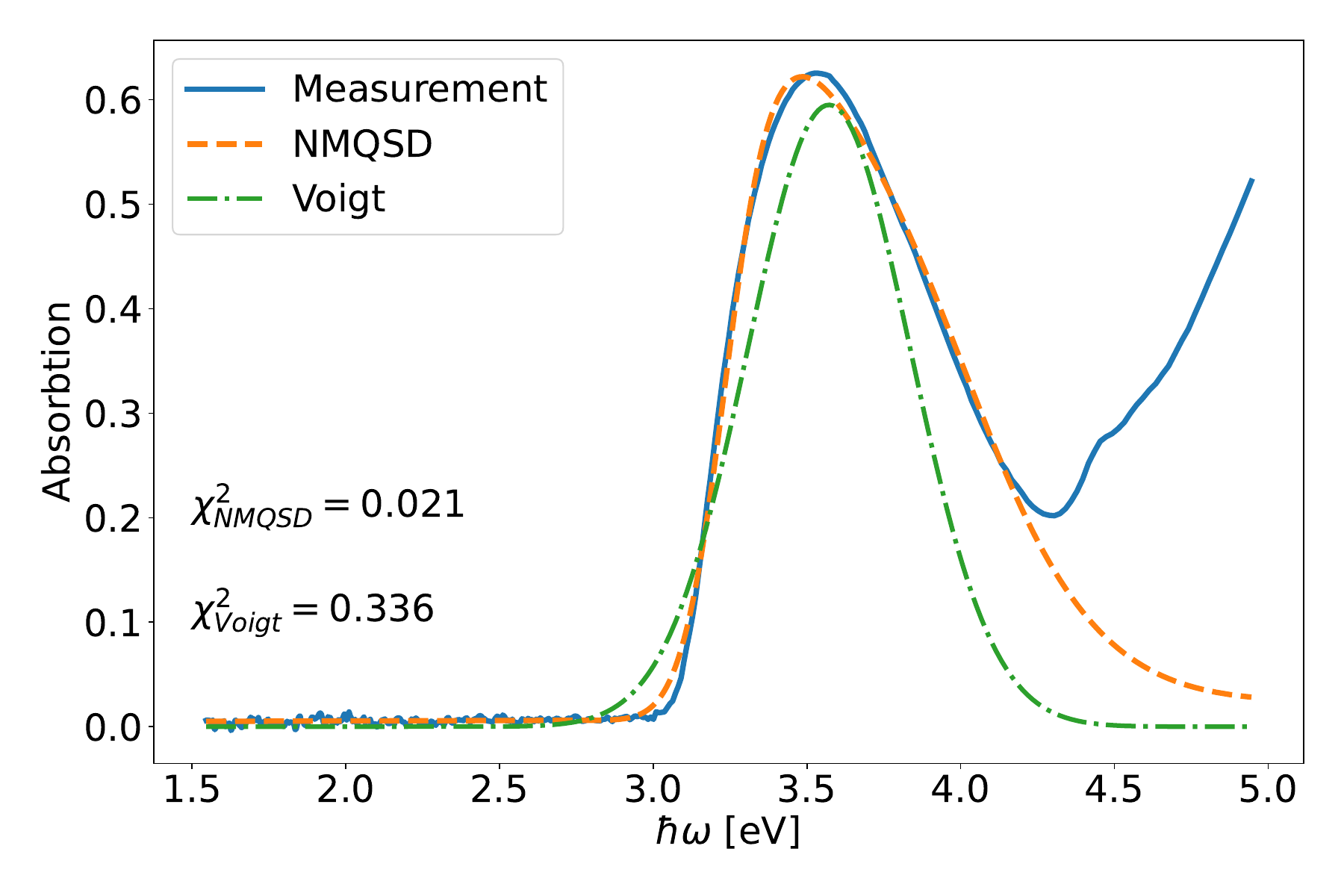}
  \caption{\label{fig:single_molecule_abs} Absorption of a 60 nm thick thin film of TDAF molecules. The exciton energy is approximately $3.6$ eV. Our model (NMQSD) fits well to the measured data, whereas a blind fit with a Voigt lineshape performs weaker in line with reported $\chi^2$ values for each fit.The errorbars in the fit are smaller than the linewidth.}
\end{figure}

\paragraph{Description.}
The exciton  of the TDAF is determined from experimentally measured absorption of a 60-nm-thick film of TDAF on a quartz substrate. The absorption is defined as $A=1-T-R$, where $T,R$ are the fractions of the transmitted and reflected light, respectively. To accomplish the measurement of the reflected and transmitted light from the film without increasing the optical path length, the sample was excited at a $15^\circ$ angle. We will model this process by computing the refractive index from a microscopic model and then the reflected and transmitted light using TMM~\cite{tmm_byrnes}.
\paragraph{Model.}
The dynamics of the molecule is governed
by the Holstein model
\begin{align}\label{eq:Holstein_model}
  H_H = (\varepsilon_S+\zeta) \sigma_+\sigma_- +\sum_\lambda\omega_\lambda b_\lambda^\dagger b_\lambda
  +\sigma_+\sigma_-\sum_\lambda g_\lambda (b_\lambda+b_\lambda^\dagger).
\end{align}
We denote by $K=\sigma_+\sigma_-$ from now on. 
The system Hamiltonian, and the coupling operator are in this case
\begin{align}
  H_S = (\varepsilon_S+\chi)K,\quad  L =K=K^\dagger,
\end{align}
where $\zeta$ is a disorder parameter. The coupling operator is hermitian.
The NMQSD equation in this case is
\begin{align}\label{eq:NMQSD_single_molecule}
  \frac{\diff}{\diff t}\ket{\psi_t}
  =& \left(-i H_S + z_t^*K+\xi_t^*\sigma_--\frac{\gamma}{2}K\right)\ket{\psi_t}\notag\\
  &-K\int_0^t\diff s\, \alpha(t-s)\frac{\delta}{\delta z_s^*}\ket{\psi_t},
\end{align}
where $\alpha(t-s)$ is the thermal BCF.
We model the radiative damping by and additional white noise process $\xi_t^*$ with correlation
$\mean{\xi_t\xi_s^*}=\gamma\delta(t-s)$. For computing the susceptibility we can set both noise
terms to zero.
The spectral density is taken to be superohmic with exponential cut-off~\cite{Martinez-Martinez2019}
\begin{align}\label{eq:superohmic}
    J(\omega) = a\frac{\omega^u}{\xi^{u-1}}e^{-\omega/\xi},\quad u=3,
\end{align}
where $a$ is parameter additionally controlling the coupling strength.
In the limit that the radiative damping is small compared to 
other parameters of the system the model admits a solution
\begin{align}\label{eq:single_molecule_solution}
    \ket{\psi_t}=\exp\left(\left[-i(\varepsilon_s+\zeta-i\gamma/2)t-g(t)\right]K\right)\ket{\psi_0},
\end{align}
where $g(t)=\int_{-\infty}^t\diff s\,\int_{\infty}^s\diff s\,\alpha(s)$. We set the integration limits in this way to remove boundary terms and the reorganization energy term, which we absorb to the singlet energy. We do not highlight explicitly in the notation that the noises $\xi_t^*=z_t^*=0$ anymore. The disorder $\zeta$ is distributed according to a Gaussian distribution with zero mean and standard deviation $\sigma$. 
\paragraph{Susceptibility.}
The dipole operator for this system is 
\begin{align}\label{eq:mol_mu}
    \mu = \mu(\sigma_++\sigma_-),
\end{align}
with the abuse of notation we use the same symbol for the transition dipole operator and the transition dipole moment. The initial state to use is the ground state of the molecule $\ket{g}$. Inerting this ground state and  computing the average over the disorder gives the following expression for the dipole correlation function
\begin{align}\label{eq:lineshape_function}
    M(t) = \exp\left(-i(\varepsilon_s-\frac{1}{2}\sigma^2t^2-i\gamma/2)t-g(t)\right)
\end{align}
In the case that $g(t)=0$ this corresponds to the Voigt lineshape. The susceptibility is obtained by taking the Laplace transform from the dipole correlation function~\eqref{eq:lineshape_function} and the refractive index can be readily computed.
\paragraph{Thin film absorption.}
The first singlet excited state is of the system is at $\varepsilon_S\approx 3.6$ eV as can be seen from the experimental trace in
Fig.~\ref{fig:single_molecule_abs}. The radiative lifetime of the TDAF thin film is reported to be 133 ps ($\sim 10^{-5}$ eV)~\cite{Toffanin2010}. 
When fitting the model to the data we ensure that the standard deviation of the disorder parameter is larger than the radiative life time so that the solution~\eqref{eq:single_molecule_solution} remains valid.
For the above mentioned parameter values the thermal contributions to the dipole correlation function are insignificant and we use the zero temperature BCF when fitting the model to the experimental data.
 
The parameter values found in the fitting process are $\varepsilon_s=3.6$ eV, $\sigma=0.14$ eV, and $\xi = 0.09$.We kept the coupling strength parameter at a fixed value $a=1$ and $\gamma=5\times 10^{-5}$ eV and included a background refractive index $n_{bg} = \sqrt{1.5+0.015i}$ to model the residual absorption at small energies. The reorganization energy for the fitted parameters is $\lambda_s \approx 0.19$ eV which is agreement with the values reported in the DFT calculations~\cite{abdelmagid2023}. The reorganization energy is absorbed in $\varepsilon_s$. The measured data, the fit using our model and a blind fit to a Voigt lineshape are shown in Fig.~\ref{fig:single_molecule_abs}. The $\chi^2_{NMQSD}$ is significantly smaller than the $\chi^2_{Voigt}$. We obtain the Voigt lineshape by neglecting the molecular vibrations contained in the term $g(t)$ in Eq.~\eqref{eq:lineshape_function}. This term is responsible for asymmetric broadening of the lineshape on higher energies.

\paragraph{Summary.} We model the thin film of TDAF molecules as two level systems which each are coupled to their respective molecular vibrations and are probed by a weak classical electromagnetic 
field. We assume that there is no spatial disorder but the energies of the excitons are distributed according to Gaussian distribution with mean $\varepsilon_s$ and variance $\sigma$. The macroscopic polarization is then obtained by multiplying the induced dipole moment with the number of molecules in the sample.  The inclusion of the disorder is motivated by the likeness of the experimental absorption lineshape to a Voigt profile and the fact that we found that thermal effects in our model were negligible in this parameter regime. The asymmetry in the absorption lineshape is explained by the coupling to the vibrational bath with spectral density given in Eq.~\eqref{eq:superohmic}. We find the parameter values by
fitting the model to the experimentally measured absorption using TMM. The fitted parameters are the 
excitonic energy ($\varepsilon_s$), energy disorder ($\sigma$), coupling strength $a$ and the cut-off $\xi$.
We keep the radiative lifetime of the exciton and the background refractive index constant during the fit.
The fit of our model is very good and the obtained parameter values are in agreement with what is obtained from DFT calculations~\cite{abdelmagid2023}.
\section{Microcavity polariton reflectivity}\label{sec:polariton_reflectivity}
\paragraph{System.}
The system is a 80 nm TDAF film in a cavity formed by two aluminium mirrors with thicknesses 25 nm and 100 nm. The reflectivity of this system can be measured experimentally as a function of the angle and energy of the incoming light. We again compute the refractive index from a microscopic model and then compare the reflectivity calculated using TMM with the experimental data.
\paragraph{Model.}
We follow~\cite{Michetti_2005,Tichauer2021} in the construction of the model.
The system consists of $N$ molecules in a planar microcavity.
The Hamiltonian for the molecules is 
\begin{align}
    H_M = \sum_{m=1}^N \varepsilon_S K_m.
\end{align}
Photons in the cavity have the energy
\begin{align}
    \omega(\vk)=\frac{c}{n_r}\sqrt{k_x^2+k_z^2},
\end{align}
where $c$ is the speed of
light in the vacuum and $n_r$ is the refractive index of the propagation medium. $\vk$ is the wave vector of the light which is assumed to be in the $x$-$z$ plane. Mirrors at $x=\pm\frac{L}{2}$ confine the light in
the $x$-direction so that the wavevector $k_x =m\pi/L$, where
$m\in\integer$. The cavity frequency at zero incidence is
$\omega_0=m\pi c/(n_r L)$. We restrict ourselves to $m=1$ case, so
that the cavity dispersion relation is
\begin{align}\label{eq:cavity_dispersion}
    \omega(k_z) = \frac{c}{n_r}\sqrt{(\pi/L)^2+k_z^2}.
\end{align}
Each mode has two orthogonal transverse polarizations $\vu_1$ and
$\vu_2=\frac{\vk\times\vu_1}{\abs{\vk}}$. 
The Hamiltonian for the cavity modes is then
\begin{align}
    H_C = \sum_{k_z} \omega(k_z)a_{k_z}^{\dagger}a_{k_z},
\end{align}
where $a_{k_z}$ is the annihilation operator for the cavity mode with the wave vector $z$-component $k_z$.
The cavity modes couple to the transition dipole moment $\vmu_j$ with coupling strengths
\begin{align}
    g_m(k_z) = -\vmu_m\cdot\vu\sqrt{\frac{\omega(k_z)}{2\varepsilon_0 V}},
\end{align}
where $\vu$ indicates the direction of the electric field of the
confined mode, $\varepsilon_0$ is the vacuum
permittivity and $V$ the cavity mode volume. We assume that the polarization of the cavity modes $\vu$, the dipole moments $\vmu_j$ and the polarization of any incoming light are all aligned in the $y$-direction. Considering a system with polarization parallel to the applied electric field, see Eq.~\eqref{eq:linear_polarization}.
The coupling between the molecules and the cavity modes is given by the
Tavis-Cummings interaction
\begin{align}
    H_{MC} = \sum_{m=1}^N \sum_{k_z} g_m(k_z)\left(\sigma^+_m a_{k_z}e^{ik_z z_m} +h.c.\right),
\end{align}
where $e^{\pm ik_z z_j}$ describes the phase of the electric field of the cavity mode at the position $z_j$ of the molecule $j$.

The coupling of each molecule to local vibrational modes is the same as in Sec.~\ref{sec:thin_film}
\begin{align}
  H_{ME} = \sum_{m=1}^N \sum_\lambda g_\lambda K_m(b_{\lambda,m}+b_{\lambda,m}^\dagger),
\end{align}
where we assume that each molecule couples to its own bath of vibrational modes.
The total Hamiltonian for the system is then
\begin{align}\label{eq:HTC}
  H_{HTC} = H_M + H_C+H_{MC}+H_{ME}+H_E,
\end{align}
where $H_E$ is the free Hamiltonian for the vibrational modes
\begin{align}
  H_E = \sum_{m=1}^N \sum_\lambda \omega_\lambda b_{\lambda,m}^\dagger b_{\lambda,m}.
\end{align}
In addition the cavity modes are damped with a 
rate that does not depend on $k_z$ and denote it by $\kappa$. The cavity damping
is described in terms of additional zero mean white noise processes $w_{t,k_z}$
with correlations
\begin{align}\label{eq:cavity_damping}
  \mean{w_{t,k_z}w_{s,k_z'}^*}=\kappa\delta_{k_z,k_z'}\delta(t-s),\quad
  \mean{w_{t,k_z}w_{s,k_z'}^*}=0.
\end{align}
Similarly, the excitons are damped with the rate $\gamma$, which is described by white noise processes $\xi_{t,j}$. 
The evolution of the system is given by the NMQSD equation which is trivially extended from the ones given earlier in the manuscript. Namely, we read off the terms of the NMQSD equation from the HTC Hamiltonian (Eq.~\eqref{eq:HTC}), add independent excitonic and cavity dampings described by the white noise terms (Eq.~\eqref{eq:cavity_damping}). Differently to the previous case, the thermal effects are significant and we use the non-zero temperature bath correlation function (Eq.~\eqref{eq:BCF_finite_temp}) in the subsequent calculations.

\begin{figure*}[t]
  \includegraphics[width=0.99\textwidth]{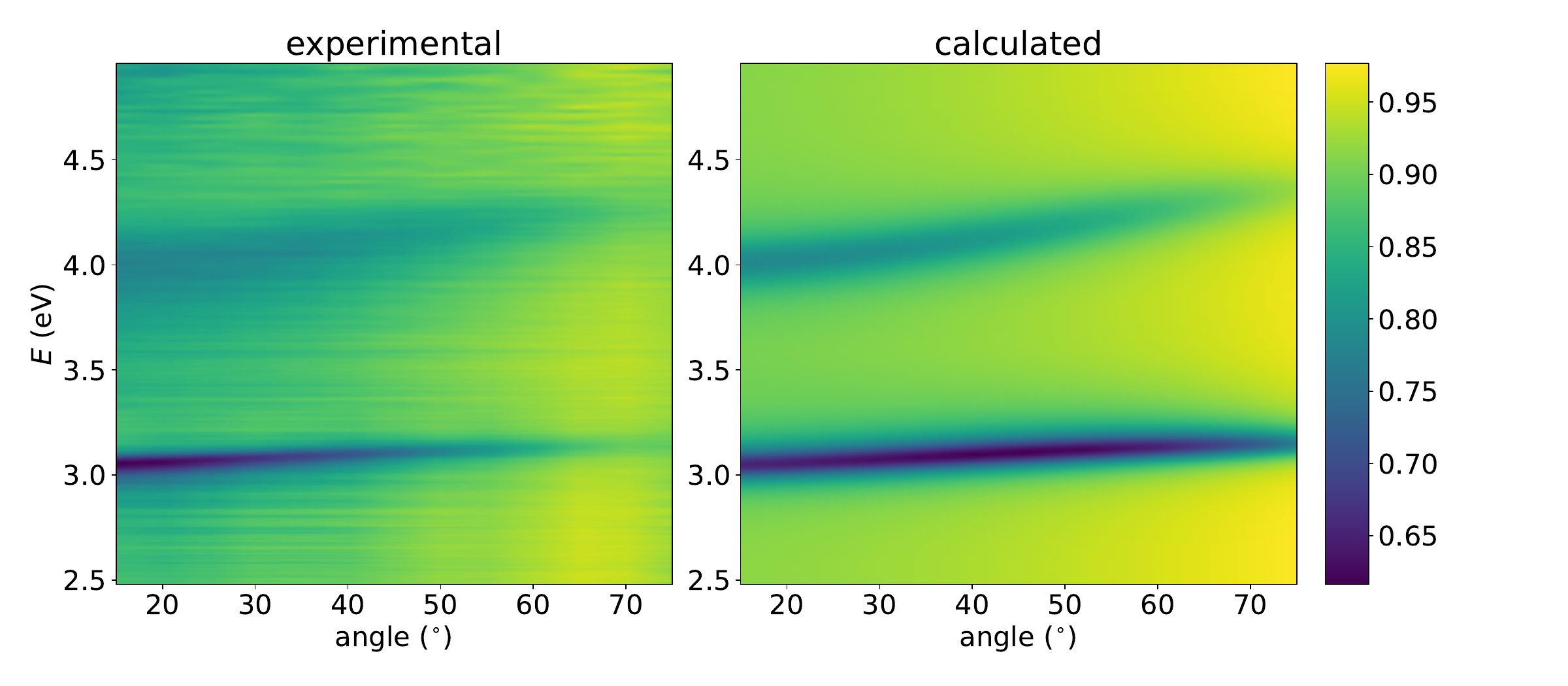}
  \caption{\label{fig:reflectivity_heatmap} Experimental and calculated reflectivities for the cavity system. The qualitative agreement of the experiment and the theory is good. The reduced reflectivity of the upper polariton is due to the coupling to the vibrational modes. The agreement between our theory and the experiment is better at smaller angles.}
\end{figure*}

\begin{figure*}[t]
    \includegraphics[width=0.99\textwidth]{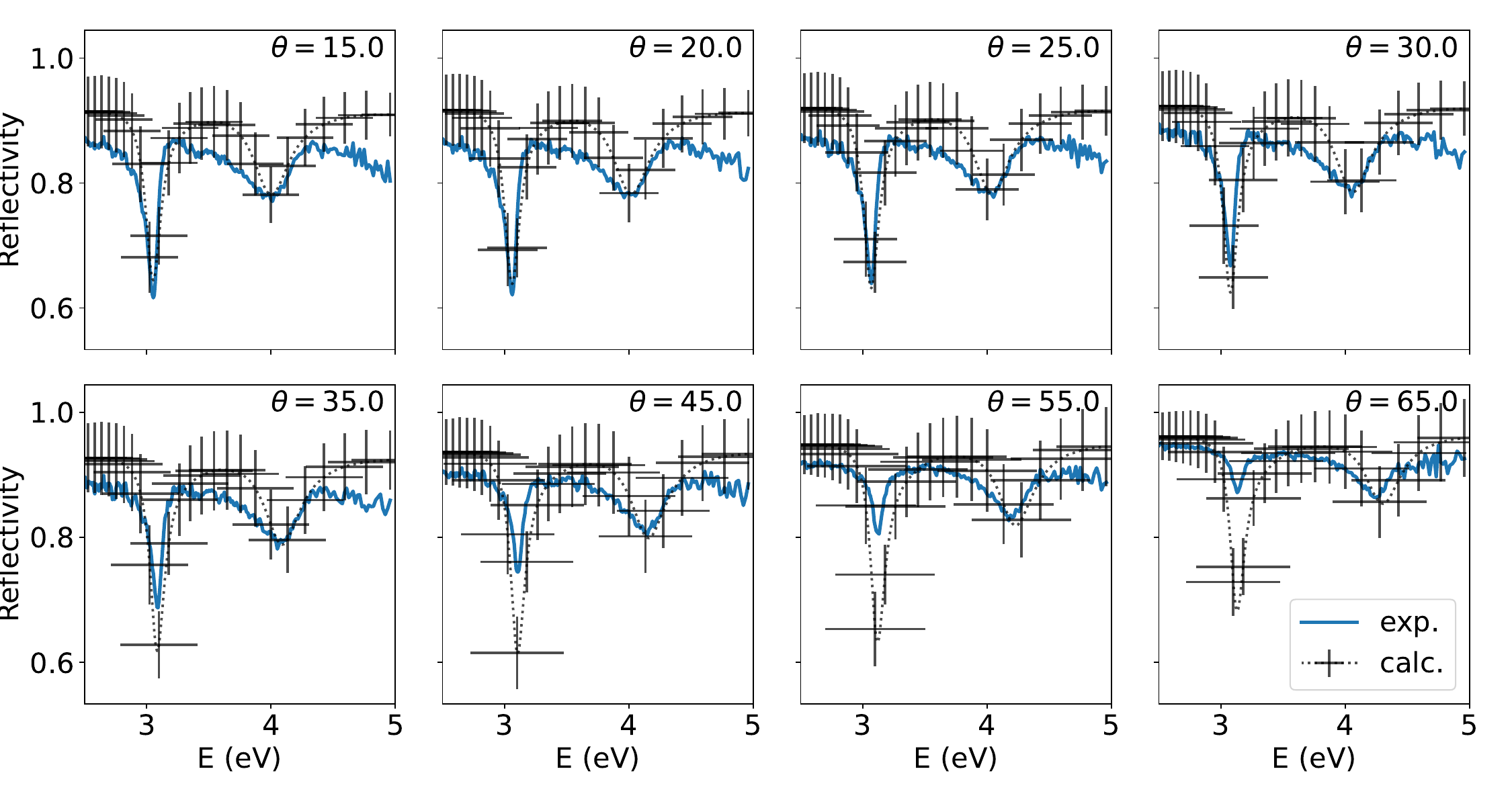}
    \caption{\label{fig:polariton_model_error2} Error estimates for the polariton model. For small angles $\theta\leq 30^\circ$ the location and amplitude of the computed and measured upper and lower polariton peaks are within the estimated error bars. For larger angles the model is in good qualitative agreement with the measured reflectivity and predicts the locations of the polariton peaks.}
\end{figure*}

\paragraph{Susceptibility.}
We write the transition dipole moments for the molecules similarly as in Eq.~\eqref{eq:mol_mu} for each individual molecule.
The dipole correlation function of this system can be calculated using \eqref{eq:dipole_correlation_NMQSD}. In this case the initial state \eqref{eq:initial_condition}
\begin{align}\label{eq:initial_polariton}
  \ket{\psi_0} = \frac{1}{\mu_{tot}}\sum_{m=1}^{N+1} e^{ik_zz_m}\mu_m\ket{g},
\end{align}
is used. We take into account the possibility of light being absorbed by the cavity mode by an additional dipole moment $\mu_{N+1}(k_z) = \mu_c a_{k_z} + \mu_c^* a_{k_z}^\dagger$ with $z_{N+1}=0$ reflecting the fact that the cavity mode is delocalized in the whole cavity volume. We can then calculate the linear susceptibility of the system from equation \eqref{eq:linear_susceptibility} and the refractive index. 
We use the refractive index to calculate the reflectivity of the cavity system by using TMM for incoming light with the angular frequency $\omega$ and $z$-component of the wave vector $k_z$. These are related to the angle $\theta$ of the incoming light by
\begin{align}
    \omega = \frac{ c \abs{\vk}}{n_r}= \frac{ c k_z}{n_r \sin(\theta_c)}= \frac{ c k_z}{\sin(\theta)},
\end{align}
where $\theta_c$ is the angle inside the cavity and we have used the Snell's law. 
\paragraph{Reflectivity.}
 We show the computed and measured reflectivity in Fig.~\ref{fig:reflectivity_heatmap} as a density plot. The quantitative agreement is good at small angles. For larger angles the locations of the polariton modes are in good qualitative agreement. In Fig.~\ref{fig:polariton_model_error2} we show the computed and measured reflectivity as a function of the energy for different angles. There the disagreement at larger angles is more prominently visible. In Fig.~\ref{fig:polariton_model_error2} we also present the estimated errorbars. We can conclude that for angles up to approximately $30^\circ$ the model and the experiment are within the estimated errorbars. We discuss the errors involved in Sec.~\ref{sec:conclusions_and_outlook}. In the TMM calculations the front mirror and the TDAF film is replaced with a material that has the computed refractive index. We use the parameters found in the fitting process in Sec.~\ref{sec:thin_film}. However, we set molecular disorder to zero and discuss this point in Sec.~\ref{sec:conclusions_and_outlook}. The number of molecules $N=30$ and the number of cavity modes is 21. The cavity decay rate $\kappa=0.21$ eV is estimated from experimental photoluminescence of the cavity system.
We fitted the positions and depths of the reflectivity minima to the experimental data and got the values $n_r=2$ for the refractive index that determines the cavity dispersion relation \eqref{eq:cavity_dispersion}, $E(0)=3.42$ eV for the energy of the cavity mode with $k_z=0$, $\Omega=0.92$ eV for the Rabi splitting, and $\mu_c = 2\mu_j$ for the magnitude of the cavity dipole moment.
At small angles the lower polariton is mostly photonic and the upper polariton is mostly excitonic. Without vibrational coupling and cavity mode absorption this would lead to pronounced absorption of the upper polariton in comparison to what is observed experimentally. Including quantum mechanical coupling to the vibrational modes and taking in account the possibility for cavity mode absorption leads to an agreement between the experiment and our theory.
If the cavity mode absorption is not included in \eqref{eq:initial_polariton}, the upper and lower polaritons absorb roughly the same amount, which is not what we see in the experiment. 

\paragraph{Summary.} We extend the thin film model from Sec.~\ref{sec:thin_film} by including the dispersive cavity mode and spatial locations of the molecules. We have excluded the energy disorder of the excitons from the model. This is done because the effect of the energy disorder led to poor agreement with the model and the data. It may be that the approximations we do in our modeling do not correctly take into account the disorder and the coupling to the cavity modes. On the other hand, neglecting the effects of the exciton broadening are in line with the observations~\cite{Wurdack2021,pandya2022tuning}, where the coupling to the cavity mode diminishes the exciton broadening effect. There, however, the cavities had higher Q factors as in our case. We estimate the error in our modeling by Monte Carlo sampling. We sample the computed reflectivity with randomly choosing the input parameters. We assume that all of the deviations are independent and distributed normally around the parameter values used in Fig.~\ref{fig:reflectivity_heatmap}. We consider the following deviations: The input energy has a standard deviation $\sigma_E=10^{-6}$ eV, the input angle $\sigma_\theta = 0.001^\circ$ and the thickness of the system $\sigma_d=10^{-9}$ m. We assume that the computed refractive index has energy dependent standard deviation $\sigma_n = 10^{-2}e^{-(\omega-\omega_0)/2}$ for the real and imaginary part, where $\omega_0$ is the smallest energy used. The deviation is smaller for higher energies, which necessary for obtaining meaningful errorbars around the polariton energies. The error bars in the location of the polaritons (x-axis) is taken to be the average variance of the estimated variances of the upper and lower polariton peak locations.

\section{Conclusions and outlook}\label{sec:conclusions_and_outlook}
We have computed the susceptibility for organic microcavity polaritons from the Holstein-Tavis-Cummings model using non-Markovian Quantum State diffusion. This approach is very efficient since we can compute susceptibility from evolving just a single state vector, so that density matrix computations are not needed. We have shown that our  model can explain the absorptive properties of TDAF thin films and the reflectivity of the TDAF microcavity polaritons. 
In the thin film case the model explains the asymmetric shape of the absorption line very well. 
In the polariton case we do have good quantitative agreement around the polariton peaks at small angles ($\leq 30^\circ$) and good qualitative agreement at larger angles. We verify the modeling results with estimated error bars. 

We have identified several open questions that will be the focus of our forthcoming investigations. We have seen that energetic disorder leads to a good fit in the thin film case but needs to removed in microcavity polariton case: i) Possible explanations used in the literature are the cavity filtering effect~\cite{Wurdack2021,pandya2022tuning} or motional narrowing~\cite{Whittaker1996,Kavokin1997}. The former may not be the right explanation in this case since the cavities used in this work have such poor Q factors. The latter explanation fails as the spectroscopy we use conserves the planar wave vector of the cavity~\cite{kavokin2007microcavities}.  

ii) In this work we have relied on perturbative solutions to the NMQSD equation as they provided reasonable fits to the experiment. It will be interesting to investigate situations where such perturbative approaches fail. In such cases, there may be more complex intramolecular dynamics which may involve also the spin orbit coupling between the singlet and triplet states~\cite{Martinez-Martinez2019}. 

iii) Lastly, we point out that state vector based approaches, such as NMQSD, open up a new way to study delocalization degree of the polaritons and polariton transport as it is possible to observe dynamically how the localization due to coupling to vibrational degrees of freedom and delocalization due to cavity coupling compete.



\section{Methods}\label{sec:methods}
\subsection{Fabrication}

The samples were fabricated using thermal evaporation at a base pressure below $10^{-7}$ Torr (Angstrom Engineering physical vapor deposition system). We used 15$\times$15 mm$^{2}$ quartz substrates that were cleaned by sonication for 10 min in soapy water (3 $\%$ Decon 90), acetone, and isopropanol, respectively and dried with nitrogen. A 100-nm-thick aluminium was deposited on top of the substrate as the bottom mirror, followed by the deposition of 80~nm TDAF as the active layer, 1~nm of LiF, and a 25-nm-thick aluminium layer as a top polariton microcavity mirror.

\subsection{Characterization}
The TDAF absorption and polariton angle-resolved reflectivity were measured with a spectroscopic ellipsometer (J.A. Woollam VASE) in reflectivity and transmission configuration. To extract the absorption of TDAF film, we measured transmitted and reflected light at a $15^\circ$ excitation angle, which represents the minimum angle our setup can measure reflectivity and adds only a 2$\%$ increase in the optical path.

\acknowledgments
This project has received funding from the European Research Council under the European Union's Horizon 2020 research and innovation programme (grant agreement No. [948260]) and from Business Finland project Turku-R2B-Bragg WOLED with decision number 1951/31/2021. K.L. would like to thank Walter Strunz, Valentin Link, Kai Müller and Christian Schäfer for fruitful discussions.
\bibliography{refs}

\end{document}